\documentstyle[12pt]{article}
\begin{document}
\bibliographystyle{unsrt}
\setlength{\baselineskip}{0.4in}
\begin{center}
{\Large  Nuclear Spin-Lattice Relaxation Induced by Thermally}
{\Large   Fluctuating Flux Lines in Type-II Superconductors}
\end{center}

\vskip .15in
\begin{center}
Lei Xing and  Yia-Chung Chang
\vskip 0.7cm

{\sl  Department of Physics and Materials Research Laboratory\\
University of Illinois at  Urbana-Champaign,  Urbana, IL 61801}
\vskip 0.7cm

\end{center}

\textwidth =6.9in

\vskip .2in
\normalsize
Thermal motion of the flux lines (FL)  gives rise to
fluctuating magnetic fields. These dynamic fields couple to the nuclei
in the sample and relax the nuclear spins.
Based on a model of harmonic  fluctuations, we provide
a theoretical description  of  the nuclear spin-lattice
relaxation (NSLR) process due to the fluctuating FLs in clean
type-II superconductors.
At low fields, the calculated longitudinal  relaxation rate $T_1^{-1}$
is enormously enhanced at temperatures just below $T_c$.
At intermediate fields,  the resulting $T_1^{-1}$ exhibits
a peak structure as a  function of temperature, which is eventually
suppressed  as the field is increased. The vibrational
modes which have components propagating along the FLs
play an essential role in  the  $T_1^{-1}$  process.

{\hfill January, 1994}

\vskip 3cm
\noindent
To appear in Phys. Rev. Lett.
\newpage
Several recent experiments and theories have pointed to the importance
of thermal fluctuations in the thermodynamics  of  type-II
superconductors.\cite{gammel,nelson,brandt}
These fluctuations also have pronounced effects on
the dynamic properties of the system.  Recent measurements
of the NSLR  rate $T_1^{-1}$ in the organic superconductor
 $\kappa$-(ET)$_2$Cu[N(CN)$_2$]Br (ET) by De Soto and Slichter et al.
provide an excellent  example of this.\cite{desoto}
It was shown that there is an enormous  enhancement of the relaxation rate
of $^1$H nuclei in the superconducting state relative
to the normal state Korringa value. It is believed that the
experimentally observed
peak structure does not belong to the category of
the Hebel-Slichter coherence peak,
rather it is caused by  the thermal motion of
the fluxoid system.\cite{desoto}

In this letter, we attempt to provide a theoretical description of  NSLR
in clean type-II  superconductors induced by the
fluctuating magnetic fields of the flux line lattice (FLL).
In the presence of fluctuating FLs,  the local magnetic
field at nuclear  sites varies in time, which  may
contribute to the NSLR.  This  mechanism
becomes dominant in a system such as   ET
when  relaxation through other channels (e.g.
through the interaction with  conduction electrons)
is suppressed.\cite{desoto} We will confine our study to the
region $H \ll H_{c2}$, where the mean distance  $a_0$ between FLs
is much larger than the radius of the vortex core $\xi$ (coherence length).
Since  we aim to give a general treatment of the  phenomenon,
we  will, for simplicity,  consider a model   of  continuous FLs based on the
3D London description.
The case of layered superconductors  with the external magnetic field
 perpendicular to the layers (chosen along the z-direction)
 could be handled similarly.

We specify the axis of the $j$-th FL at time $\tau$ by a 3D vector
$\vec r_j(\tau)$.
The local  field $\vec B(\vec r,\tau)$ varies with time $\tau$ only
through the motion  of the FL
and is given by\cite{brandt}
\begin{equation}
{\vec B}(\vec r,\tau) = {\frac {\phi_0}{4 \pi  \lambda^2}}
\sum_j \int d \vec s_j
\frac {\exp(-|\vec r - \vec r_j(\tau)|/\lambda)}{|\vec r - \vec r_j(\tau)|} =
\frac {V}{(2 \pi)^3} \int d^3 \vec k  \;{\vec b}(\vec k)  \frac{1}{N L}
\sum_{j} \int d z_j e^{ i \vec k \cdot (\vec r - \vec r_{j}(\tau))}~~,
\end{equation}
where $\vec b (\vec k) = B ( {\hat z} k_{\perp}^2 - {\vec k}_{\perp} k_z)/
k_{\perp}^2 (1 + k^2 \lambda^2)$, $\vec k = (\vec k_{\perp}, k_z)$
and  $\vec k_{\perp} = (k_x,k_y)$, $\lambda$ is the magnetic penetration depth,
the line integral $\int d \vec s_j$ is along the $j$-th FL, $V$ is
the volume of the sample, and $N$ and  $L$ are the number and the length of
FL respectively.
To obtain the Fourier transform of $\vec B(\vec r,\tau)$ we
used $\nabla \cdot \vec B = 0$. A similar expression for the layered system
can be found in Refs. \cite{brandt,song,levnmr}.

It is seen clearly from Eq. (1) that the thermal motions of FLs lead
to fluctuating magnetic fields. These fields are coupled  to  the nuclear
moments in the sample. This coupling is described
by the interaction ${\cal H}(\tau)
= -\gamma_n \hbar \sum_n    \vec I \cdot \vec B(\vec r_n,\tau)$,
where $\gamma_n$ is the nuclear gyromagnetic ratio,
$\vec I$ is nuclear spin operator, and $\vec r_n$ is
the position of the $n$-th nucleus. We then have\cite{slichter}
\begin{equation}
\frac{1}{T_1} = 2 \gamma_n^2 \sum_{\alpha=x,y}
| (\uparrow |I_\alpha|\downarrow) |^2 \int {\frac {d \tau}{2 \pi} }
\cos \omega_0 \tau   {\langle{\overline {B_\alpha (\vec r,0) B_{\alpha}
(\vec r,\tau)}}\rangle} ~~,
\end{equation}
where  $\omega_0=\gamma_n B$,
$\langle ~~  \rangle$ represents the  thermal average  of the
FL configurations. Here we have taken an average over the relaxation rates
of all the  nuclei (denoted  by   the horizontal bar in Eq. (2)),
which will be  evaluated by spatial integration.
Using Eq. (1) we  recast  Eq. (2) into
\begin{equation}
\frac{1}{T_1} = \gamma^2_n K(\omega_0,B) =\gamma_n^2 \frac{V }{(2 \pi)^3}
\int d^3 \vec k
|\vec b_\perp(\vec k)|^2 \frac{1}{NL}  {\tilde S}(\vec k,\omega_0)~~,
\end{equation}
where $K(\omega_0,B)$ is the fluctuation spectral density function,
${\tilde S}(\vec k,\omega_0) = [S(\vec k,\omega_0)+
 S(\vec k,-\omega_0)]$, and
\begin{equation}
S(\vec k, \pm \omega_0)= \int_{-\infty}^{+\infty}
\frac{ d \tau}{2 \pi}  e^{ \pm i\omega_0 \tau} \frac {1}{N L}
\sum_{ j j'} \int d z_j \int dz_{j'}
\langle e^{i \vec k \cdot
\vec r_{j}(0)} e^{- i\vec k \cdot \vec r_{j'}(\tau)} \rangle
\end{equation}
is the dynamic structure factor of the FLs, which is entirely determined
by the FL structure and corresponding time evolution spectrum,
without reference to any properties
of the nuclei. We emphasize that  our result (3) is quite general,
applying  even to the flux liquid and glass states. In the following we will
focus our study on  the FLL state.

The problem of finding $T_1^{-1}$ is now reduced to obtaining
the dynamic structure factor $S(\vec k,\omega_0)$ of  the FLs,
which is a fundamental dynamic quantity
and of importance  in many other applications (e.g. the inelastic
scattering). Evaluation of $S(\vec k,\omega_0)$
requires information about the dynamics of the FL  fluctuations.
 Bulaevskii et al.\cite{levnmr} have used
the Langevin equations for the overdamped motion of pancakes
in interpreting their $T_1^{-1}$ data
in Tl$_2$Ba$_2$CuO$_6$,
and found a monotonic  decrease  of $T_1^{-1}$ with  temperature.
At this time  the dynamics of the FLs are much less well
understood than their static  properties.  A clean translational
invariant superconductor is thus a valuable model system which can
provide significant insight into the physics involved.
In this case it is clear that the vortex motion
is reversible.  Once this fundamental case is  understood,
various complications (e.g. including damping of the vortex motion)
can be introduced.

For a clean superconductor,
Fetter has derived a set of dynamic equations for the FLs by using
the London model, which  provides a basis for our study.\cite{fetter}
We assume that the deformation of  a rectilinear FL is  small
so that the vortex axis can be  specified by
$\vec r_i(\tau)=(\vec R_i^{(e)} + \vec u_i (z,\tau),z)$,
where $\vec u_i(z,\tau)$ is the displacement of
the $i$-th FL from its equilibrium
position $(\vec R_i^{(e)},z)$.
In the harmonic approximation to the total energy of the FLL,
the motion of $\vec u_i(z,\tau)$ can be
quantized by interpreting $u_{xi}(z)$ and $u_{yi}(z)$ as quantum
mechanical conjugate variables.\cite{fetter} This leads to
\begin{eqnarray}
u_{xi}(z,\tau)&=&\frac{1}{\sqrt{\rho {\overline \kappa} L N  }}
\sum_{\vec q_{\perp}
q_z} \sqrt{\frac{\hbar  {\tilde \Omega}} {2 \omega} }
( a_{- \vec q_{\perp} -q_z} e^{-i\omega \tau} +
a_{\vec q_{\perp} q_z}^{\dagger} e^{i\omega \tau} )
e^{i \vec q_{\perp}
\cdot \vec R_i^{(e)} +i q_z z}~~,\\
u_{yi}(z,\tau)&=&\frac{-1}{\sqrt{\rho {\overline \kappa} L N }}
\sum_{\vec q_{\perp}
q_z} \sqrt{\frac{\hbar } {2 \omega {\tilde \Omega} }}
[({\overline \alpha} +i \omega) a_{-\vec q_{\perp} -q_z}
e^{-i\omega \tau} +
({\overline \alpha} - i \omega) a_{\vec q_{\perp} q_z}^{\dagger}
e^{i\omega \tau}] e^{i \vec q_{\perp}
\cdot \vec R_i^{(e)} +i q_z z}~~,
\end{eqnarray}
where  ${\overline \kappa} =h/2m$,
$ {\tilde \Omega} = {\overline \Omega} + \frac {1-q_z^2 \lambda^2}
{1+q_z^2 \lambda^2} {\overline \eta} + {\overline \xi}$,  $\rho$ is
the mass density of the superelectrons, and the summation over $\vec q_\perp$
is restricted to the first Brillouin zone. In the above equations,
${\overline \Omega}$,
${\overline \alpha}$, ${\overline \eta}$, and ${\overline \xi}$
are functions of $(\vec q_{\perp},q_z)$ and are
defined in Eqs. (43) and (44) of Ref. \cite{fetter}.  The corresponding
Hamiltonian of the FLL reads
${\cal H}_{FLL}= \frac{1}{2}  \sum_{\vec q_{\perp}
q_z}  \hbar \omega
( a_{q_{\perp} q_z}  a_{q_{\perp} q_z}^{\dagger} +
a_{q_{\perp} q_z}^{\dagger} a_{q_{\perp} q_z})$,
with the dispersion relation given by
\begin{equation}
\omega^2(\vec q_{\perp},q_z) = ({\overline \Omega} + \frac {1-q_z^2 \lambda^2}
{1+q_z^2 \lambda^2} {\overline \eta})^2 -
({\overline \alpha }^2  + {\overline \xi }^2)~.
\end{equation}
Here $a^{\dagger}_{\vec q_{\perp} q_z}$
and $a_{\vec q_{\perp} q_z}$
obey boson commutation relations, and represent the
creation and annihilation operators for a single vibrational quantum (which
will be referred  to as a ``fluxon") in the
normal mode with wave vector $(\vec q_{\perp}, q_z)$ and
frequency $\omega(\vec q_{\perp},q_z)$.

We will now derive  $S(\vec k,\omega_0)$ resulting from  Fetter's
dynamics.\cite{zhang}
As usual, we   use the Bloch identity\cite{lovesey} to write
$\langle e^{i \vec k_{\perp} \cdot
\vec u_{j}(z,0)}e^{ - i \vec k_\perp \cdot \vec u_{j'}(z',\tau)} \rangle
= e^{-2 W(\vec k_\perp)} e^{ \langle \vec k_\perp \cdot
\vec u_{j}(z,0) \; \vec k_\perp \cdot \vec u_{j'}(z',\tau) \rangle}$,
where $ W(\vec k_{\perp}) = \frac{1}{2}
\langle [k_\perp \cdot \vec u_j(z,0)]^2 \rangle$
 is the Debye-Waller factor.
The lowest order contribution arises from the one-fluxon
process, in which a nucleus flips its spin and the FLL
creates or annihilates a fluxon to conserve
the  total energy. Combining  the Bloch identity and Eqs. (4)-(6),
 and keeping only the one-fluxon term, we obtain
\begin{eqnarray}
S(\vec k,\omega_0)&=&\frac{\hbar}{2
\rho {\overline \kappa} }
e^{-2  W(\vec k_{\perp})}  \sum_{\vec q_\perp q_z}  \sum_{\vec K}
\frac{1}{\omega}  [{\tilde \Omega} k_x^2
+ {\tilde \Omega}' k_y^2 - 2 {\overline \alpha} k_x k_y ] [n \delta(
\omega + \omega_0) \delta_{k_z,-q_z}
\delta_{\vec k_\perp, -\vec q_\perp + \vec K}\\ \nonumber
&+&(1+n) \delta(\omega - \omega_0)\delta_{k_z,q_z}
\delta_{\vec k_\perp, \vec q_\perp + \vec K} ] ~~\nonumber,
\end{eqnarray}
where $\vec K$ is the reciprocal lattice  vector,
${\tilde \Omega}' = {\overline \Omega} + \frac {1-q_z^2 \lambda^2}
{1+q_z^2 \lambda^2} {\overline \eta} - {\overline \xi}$,  and $n
=\langle a_{\vec q_{\perp} q_z}^{\dagger}  a_{\vec q_{\perp} q_z} \rangle$
is the Bose-Einstein occupation factor of  fluxons.

Upon substituting Eq. (8) into Eq. (3), we obtain
\begin{eqnarray}
\frac{1}{T_1} &=& \frac{\gamma_n^2 \hbar B^2 }{2 \rho
{\overline \kappa} \lambda^2 L}
 \sum_{q_{\perp},q_z}  \sum_{\vec K} \frac{q_z^2 \lambda^2}{(\vec q_\perp +
\vec K)^2 [1+q_z^2 \lambda^2 + (\vec q_\perp + \vec K)^2
\lambda^2]^2} e^{-2W(\vec q_\perp + \vec K)}
\frac{\coth \frac{\hbar \omega}{2k_B T}}{\omega}\\ \nonumber
&\times & [{\tilde \Omega}  (q_x+K_x)^2
+ {\tilde \Omega}' (q_y+K_y)^2 - 2 {\overline \alpha} (q_x+K_x) (q_y+K_y) ]
\delta( \omega  - \omega_0)~~.
\end{eqnarray}
For arbitrary fields
and temperatures the  quantity in the summation  is  a complicated function of
$\vec K$ and $(\vec q_\perp,q_z)$,
and the summation  can only be evaluated
numerically; however, in some special cases and with some
restrictions, analytical results can be found.
The  simplest case is  a single FL in a bulk sample (this also
includes the case of well-separated FLs, i.e. $a_0 \gg \lambda$).\cite{ehren}
In this case the motion is entirely self-induced and  only
waves propagating along the FL ($\vec q_\perp = 0$) exist. Thus we have
${\overline \alpha} ={\overline \eta}
={\overline \xi} =0$.\cite{fetter} In the long wave length limit
the dispersion relation can be written as
$\omega(q_z) \approx  \frac{{\overline \kappa}}{
4 \pi} (\ln \frac{2 \lambda}{\xi} - \gamma +\frac{1}{4} )
q_z^2$,\cite{fetter,degennes}, where $\gamma =0.5772$ is  Euler's constant.
We then obtain
\begin{equation}
\frac{1}{T_1} = \frac{\hbar \phi_0 \gamma_n q_{z0}^3}
{16 \pi^2 \rho {\overline \kappa} \lambda^2} \coth \frac{\hbar \omega_0}
{2 k_B T}
\left[ \frac{1}{ 1 + q_{z0}^2 \lambda^2} + \frac{ \langle u^2 \rangle}
{\lambda ^2} e^{ \frac{ \langle u^2 \rangle} {\lambda ^2}
(1 + q_{z0}^2 \lambda^2 )}  Ei(- (1 + q_{z0}^2 \lambda^2 )
\frac{ \langle u^2 \rangle} {\lambda ^2})\right]~,
\end{equation}
where Ei$(x)$ is the exponential integral,\cite{tables}
$q_{z0}$ is determined by $\omega(q_{z0}) = \omega_0$. The dashed curve
in Fig. 1 shows $T_1^{-1}$ as a function of $t$, where
$\lambda = \lambda_0 /\sqrt{1- t^4}$,  $\langle u^2 \rangle /\lambda^2
\approx  \beta t$,\cite{fetter} $\beta= \frac{4L  k_B T_c}{\pi \rho
{\overline \kappa^2 \lambda^2} \ln (\lambda/\xi)}$ and $t=T/T_c$.
Here  we have chosen $\omega_0=5$ MHz, $\gamma_n = 2.675 \times
10^4 s^{-1} Gauss^{-1}$, $\lambda_0 =2000 \AA$, and $\beta=0.1$.

As can be seen from Eq. (9), for arbitrary vortex density, $T_1^{-1}$
is a sum of contributions from
all normal modes. It is instructive to analyze the contributions arising
from two special types of modes. First, we notice that
the modes propagating perpendicular to the FL
direction ($q_z=0$), which  corresponds to the vibrations
of the rigid FLs, do not contribute to  $T_1^{-1}$, because
there is no  transverse  field associated with these modes (as can be
seen from Eq. (2), a fluctuating
transverse field is required for longitudinal relaxation).
This indicates the essential role played by the modes
with $q_z \neq 0$  in the $T_1^{-1}$ process.
Second, consider the  vibrations propagating along the FLs
(also known as the ``helicon modes"\cite{fetter,degennes}).
In the dilute limit these modes become identical to those of a single FL.
At high fields ($a_0 \ll \lambda$),
the dispersion  relation of such modes
reads $\omega(q_z) \approx  \frac{{\overline \kappa}}{
4 \pi} (\ln \frac{2 \lambda}{\xi} - \gamma +\frac{1}{4} +
\frac{B}{H_{c1}} \ln \frac{\lambda}{\xi}  ) q_z^2$ in the long wave-length
limit.\cite{disp}
In this case we also have $K^2 \lambda^2 \gg 1$
for $\vec K \neq 0$. This allows us to
keep  the $\vec K=0$ term only in Eq. (9) and to
 obtain the contribution $ \frac{d T_{1}^{-1}}{q_\perp d q_\perp}
|_{q_\perp =0} $ from  these so-called ``helicon" modes\cite{degennes}:
\begin{equation}
\left. \frac{d T_{1}^{-1}}{q_\perp d q_\perp} \right|_{q_\perp =0}
= \frac{\hbar \phi_0 \gamma_n q_{z0}}
{16 \pi^2 \rho {\overline \kappa} \lambda^2}
\coth \frac{\hbar \omega_0}{2 k_BT}
\frac{q_{z0}^2 \lambda^2}{ (1 + q_{z0}^2 \lambda^2)^2 }~~.
\end{equation}

The NSLR rate $T_1^{-1}$
arises, in general,  from the contributions of both the ``helicon"
modes and those branches with $\vec q_\perp \neq 0$.
But one should note that not
all $\vec q_\perp \neq 0$ branches contribute
to the $T_1^{-1}$ process. This is seen
clearly from  Fig. 2, in which   we have plotted $\omega(q_\perp,q_z) /
\omega_0$  as a function of $q_z$ for a few
values of $q_\perp$. The branches with $\omega(q_\perp, q_z=0) /\omega_0
> 1$ do not contribute  to  $T_1^{-1}$ due to
the $\delta$-function  in Eq. (9).
For $^1H$ we have numerically evaluated  Eq. (9)
for $\omega_0 =70$ and $1000$ MHz
(corresponding to $B=0.37$ and $3.74$ Tesla, respectively).
The resulting $T_1^{-1}$'s  are  plotted as a function of temperature
in Fig. 1  as the solid and dotted  curves.
Notably, a relaxation peak  is found  below $T_c$. Note that
the results shown in Fig. 1  only contain the
contributions from the fluctuating FLL only.
At very high fields, the relaxation peak could be absent if $T_c(B)$
becomes less than the temperature at the peak,
or if the relaxation due to the FLL is so greatly suppressed that
it is slower than that due to other sources, e.g. the interaction with
the current carriers.

We now address the behavior of  $T_1^{-1}$ at different fields
in more detail. First of all, we  mention that
at typical NMR  frequencies (from a few  to several hundred
MHz) the condition
$ \frac{\hbar \omega_0}{k_B T_c}
\ll 1$ is satisfied for  superconductors with $T_c \sim 10 K$.
The low frequency region corresponds to  fields of a few  hundred Gauss.
The FLs are well-separated, and are very soft in this
case. $T_1^{-1}$  can be  described by
Eq. (10) and from Fig. 1 we  see that
$T_1^{-1}$ is enormously enhanced  right below $T_c(B)$.
Note that so far there seems to be  no
experimental data published  in this regime. It would be
interesting to carry out such measurements.
When the field increases to $a_0 \sim \lambda$, the interaction
between the  FLs begins to play  a role. This interaction promotes the
modes with $\vec q_\perp \neq 0$  and stiffens the FLs.  In the high
frequency range ($a_0 \ll \lambda$), only the $\vec K =0 $ term gives
important contributions to $T_1^{-1}$.  The magnitude of $T_1^{-1}$ is
greatly suppressed in comparison with the single FL case, as can be seen
in Fig. 1. A close examination of Eq. (9) suggests
that this suppression comes from   the decrease of the
thermal factor $\coth \frac{\hbar \omega_0}{2 k_B T_c}$,
the density of states of fluxons
at $\omega_0$, and the resonant wave number $q_{z0}$.
These effects  overcome the increase brought on by the prefactor
$\omega_0^2$ in Eq. (9).
One should note  that the Fetter's vortex dynamics breaks
down if the field is close to $H_{c2}$ so that vortex
fields overlap strongly. In this case the nonlocal effects
(wave vector dependence of the elastic moduli) act to
enhance the fluctuations,\cite{nelson,brandt}
and a Ginzburg-Landau type of approach
becomes relevant.

The fluctuation spectral density  function
$K(\omega,B)$ contains detailed  dynamic information of the system.
It depends on both frequency $\omega$ and magnetic field $B$.
In Fig. 3  we plot  $K(\omega,B)$ vs $\omega$ for a few $B$'s of
interest. At a fixed $B$, the function has a maximum
at a frequency $\omega_m$ which shifts upward as the field increases.
The field dependence of $K(\omega,B)$ at a given frequency $\omega$
can also be seen from Fig. 3.
At low frequencies, $K(\omega,B)$  decreases with the field,
which has been observed experimentally\cite{desoto}.
At high frequencies it increases
with $B$, and at intermediate frequencies
it has a peak structure as a function of $B$. These
are very interesting predictions which can be  tested experimentally
by measuring the NSLR rates
of different type of nuclei, or different transitions of a nucleus
with $I > 1/2$.\cite{martinedale} The measurement of $K(\omega,B)$
provides a crucial test for a vortex dynamics.

In summary, we have studied the NSLR
caused by thermally fluctuating  FLL in a clean type-II superconductor.
Its applications  for the vortex dynamics go beyond the NMR.\cite{neutron}
The NMR approach to the FLL state is  quite diagnostic,
and provides a sensitive probe to the
dynamic fluctuations of the vortices.
The simple harmonic fluctuations of the FLL seems to render a good
description of the overall features of $T_1^{-1}$ obtained in ET.
We found that the  fluctuation effects
on the $T_1^{-1}$ process is most pronounced at low fields, where the FL
is easy to ``bend" and the thermal excitations of the
fluxons with the frequency  $\omega_0$ is  effective.
In the above discussions, we have ignored the effects of damping of
the vortex motion.   We expect that inclusion of weak damping
would not change our results significantly,
and that the theory should be directly applicable to systems
with sufficiently large Hall angle\cite{dorsey}.  The harmonic
and the overdamped vortex motions (this is
believed to be the case in YBCO) represent two different important
regimes. The latter also deserves further study.
While at  present there is still no  well-accepted dynamic theory
of the vortices,
we believe  that  our calculations shed important light on   this subject.
Finally, we mention that the effects discussed
in this paper should be more pronounced
in layered superconductors since  in that  case
the FLs are more flexible.\cite{nelson}

We thank  S.M. De Soto, C.P. Slichter,  M.B. Salamon, A.J. Leggett,
G.A. Baym, L.N. Bulaevskii, R.L. Corey, R. Wortis, I. Vagner,
and A. Sudbo  for many useful discussions. We are grateful to C.P. Slichter
for communicating their experimental results prior to publication, and for
drawing our attention to the calculation of spectral density function.
Support by  the ONR grant No. N00014-89-J-1157
is also gratefully acknowledged.
\pagebreak

\pagebreak

\begin{center}
{\bf{FIGURE CAPTIONS}}
\end{center}

\noindent
FIGURE 1.  $T_1^{-1}$ as a function of $T/T_c$
at $\omega_0 =5$ MHz (dashed curve, represents
the case of a single FL),
and at $\omega_0 =70$ (solid curve) and $1000$ MHz (dotted curve).
The unit of  $T_1^{-1}$ is
chosen to be $\frac{\hbar \gamma_n \phi_0}{8 \pi^2 \rho \lambda^2 {\overline
\kappa} \lambda_0^3}$, which is about  $10^{-3}$ to $100$ Hz
for typical type-II superconductors.  Note that
we have plotted  $T_1^{-1}$ all the way up to $T/T_c=1$.
In reality, the $T_1^{-1}$ is only valid up  to $T_c(B)$.
Above $T_c(B)$, the system goes to the normal state.

\vskip 0.8cm
\noindent
FIGURE 2.  $\omega(q_z,q_\perp) /\omega_0$ at $\omega_0 =1000$ MHz
and $T/T_c=0.4$ for
$q_\perp \lambda =0, 0.15, 0.31, 0.46, 0.61,0.76$ and $0.92$ (from bottom
to top).

\vskip 0.8cm
\noindent
FIGURE 3. Fluctuation spectral density functions at $B=1000 $ (solid curve),
$3000$ (dotted curve),
and $6000$ Gauss (dashed curve).

\begin{thebibliography}{10}

\bibitem{gammel}
{P.L. Gammel, et al., {\it Phys. Rev. Lett.} {\bf 60}, 144 (1988);
{\it ibid.} {\bf 61}, 1666 (1988); {\it ibid.} {\bf 66}, 953 (1991);
D.E. Farrell et al., {\it ibid.} {\bf 67}, 1165 (1991).}

\bibitem{nelson}
{D.R. Nelson, {\it Phys. Rev. Lett.}
{\bf 60}, 1973 (1988); D.R. Nelson and H.S. Seung, {\it
Phys. Rev.} {\bf B 39}, 9153 (1989);
M.P.A. Fisher, {\it Phys. Rev. Lett.}
{\bf 66}, 3213 (1991);  D.S. Fisher, M.P.A. Fisher,
and D.A. Huse, {\it Phys. Rev.}
{\bf B 43},  130 (1991);  L.N. Bulaevskii, M. Ledvij, and V.G. Kogan,
{\it ibid.} {\bf B 46}, 366 (1992); A. Houghton et al., {\it ibid.}
{\bf B40}, 6763 (1989);
Z. Te{\v s}anovi{\' c} and L. Xing, {\it Phys. Rev. Lett.}
{\bf 67}, 2729 (1991); Z. Te{\v s}anovi{\' c}, Physica {\bf C 220},
3030 (1994);
L. Xing and Z. Te{\v s}anovi{\' c},
{\it Physica} {\bf C 196}, 241 (1992).}




\bibitem{brandt}
{E.H. Brandt, {\it Phys. Rev. Lett.}
{\bf 66}, 3213 (1991); {\it Int. J. Mod. Phys.} {\bf B 5}, 751 (1991).}


\bibitem{desoto}
{S.M. De Soto, C.P. Slichter et al., {\it Phys. Rev. Lett.}
{\bf 70}, 2956 (1993). The peak structure of $T_1^{-1}$ was also observed in
polycrystalline samples; see Refs. [7-9] of this reference for details.}


\bibitem{song}
{Y.Q. Song, W.P. Halperin et al., {\it Phys. Rev. Lett.}
{\bf 70}, 3127 (1993).}

\bibitem{levnmr}
{L.N. Bulaevskii, et al., {\it Phys. Rev. Lett.}
{\bf 71}, 1891 (1993).}

\bibitem{slichter}
{C.P. Slichter, {\it Principles of Magnetic Resonance}, Springer-Verlag,
Berlin, 1990.}



\bibitem{fetter}
{A.L. Fetter, {\it Phys. Rev.} {\bf 163}, 390 (1967).}


\bibitem{zhang}
{This dynamics was also recently employed to study the lattice state
of the vortices and antivortices in thin superconducting
and superfluid films; M. Gabay and A. Kapitulnik,
{\it Phys. Rev. Lett.} {\bf 71},
2138 (1993) and S.-C. Zhang, {\it ibid.} {\bf 71},
2142 (1993). In addition, it is worth  mentioning that a  hydrodynamic
theory of finite-mass vortices for both liquid helium and
type-II superconductors has been developed; G.A. Baym and Chandler,
Low. Temp. Phys. {\bf 50}, 57 (1983); {\it ibid.} {\bf 62}, 119 (1986);
G.A. Baym, unpublished}

\bibitem{lovesey}
{S.W. Lovesey, {\it Theory of Neutron Scattering
from Condensed Matter}, Oxford, London, 1984.}

\bibitem{ehren}
{A rough  estimate of the single FL
effect was given  in a $^1$H
NMR study of hydrated V-Ti alloy; E. Ehrenfreund et al.,
{\it Solid State Commun.}  {\bf 7},  1333 (1969).}


\bibitem{degennes}
{This  dispersion relation for a single FL
was first obtained by de Gennes and co-workers,
{\it Appl. Phys. Lett.} {\bf 2}, 119 (1963);
{\it Rev. Mod. Phys.} {\bf 36}, 45 (1964).}

\bibitem{tables}
{I.S. Grandshteyn and I.M. Ryzhik, {\it Table of Integrals, Series, and
Products}, Academic Press, New York, 1980.}

\bibitem{disp}{
In general, for $q_\perp \!=\!0$ we have
${\overline \alpha}= {\overline \eta}= {\overline \xi}= 0$
and $\omega\! =\! {\overline \Omega}$.
In  the long wave-length limit it follows that
$\omega(q_z) \approx  \frac{{\overline \kappa}}{4 \pi}
\left\{ \ln \frac{2 \lambda}{\xi} -\gamma+ \frac{1}{4} +
{\sum_i}'  \left[ K_0(\frac{R_i^{(e)}}{\lambda})+ \frac{R_i^{(e)}}{\lambda}
K_0(\frac{R_i^{(e)}}{\lambda}) \right] \right\}  q_z^2$,
where $K_0(x)$ and $K_1(x)$ are modified
Bessel functions, the primed sum means to sum over all
$\{ \vec R_i^{(e)} \}$ except $R_i^{(e)}=0$.
Compared with  $\omega(q_z)$ of a single FL, we notice
that the interaction between the FLs increases the coefficient of $q_z^2$
as the field increases.
The dispersion relation at high fields  is obtained by replacing
the lattice sum by integrals over a smooth vortex density $B/\phi_0$.}

\bibitem{martinedale}
{See J. Martinedale et al., {\it Phys. Rev.} {\bf B 47}, 9155 (1993)
and references therein.}

\bibitem{dorsey}
{Refer, for instance, A.T. Dorsey, {\it Phys. Rev.} {\bf B 46}, 8376 (1992).}

\bibitem{neutron}
{For instance, the approach can be  generalized
to study the inelastic neutron scattering  of the vortices;
L. Xing, L.N. Bulaevskii, T. Mason, and Y.C. Chang, in progress.}

\end{thebibliography}
\end{document}